\documentclass[a4paper,superscriptaddress,twocolumn,prl]{revtex4}
\usepackage{graphicx}
\usepackage[]{natbib}
\pdfoutput=1

\begin{document}

\title{A Bright Solitonic Matter-Wave Interferometer}

\author{G.D. McDonald}
\email{gordon.mcdonald@anu.edu.au}
\homepage{http://atomlaser.anu.edu.au/}
\author{C.C.N. Kuhn}
\author{K.S. Hardman}
\author{S. Bennetts}
\author{P.J. Everitt}
\author{P.A. Altin}
\author{J.E. Debs}
\author{J.D. Close}
\author{N.P. Robins}

\affiliation{Quantum Sensors and Atomlaser Lab, Department of Quantum Science, Australian National University, Canberra, 0200, Australia}

\date{\today} 

\begin{abstract}
We present the first realisation of a solitonic atom interferometer. A Bose-Einstein condensate of $1\times10^4$ atoms of rubidium-85 is loaded into a horizontal optical waveguide. Through the use of a Feshbach resonance, the $s$-wave scattering length of the $^{85}$Rb atoms is tuned to a small negative value. This attractive atomic interaction then balances the inherent matter-wave dispersion, creating a bright solitonic matter wave. A Mach-Zehnder interferometer is constructed by driving Bragg transitions with the use of an optical lattice co-linear with the waveguide. Matter wave propagation and interferometric fringe visibility are compared across a range of $s$-wave scattering values including repulsive, attractive and non-interacting values. The solitonic matter wave is found to significantly increase fringe visibility even compared with a non-interacting cloud.
\end{abstract}

\maketitle

A soliton is a single wave which propagates without dispersion. Solitons can arise in a system which is described by a weakly nonlinear dispersive partial differential equation when the nonlinear and dispersive effects cancel out. Their formation, stability and dynamics form an enormous field of rich and diverse study~\cite{RevModPhys.61.763} with applications to nonlinear optical systems~\cite{RevModPhys.83.247}, oceanography~\cite{ocean}, magnetic materials~\cite{ferromagnets,magnetoresistance}, financial markets~\cite{finance}, and biological systems~\cite{BiologicalSystems} among others.
Depending upon whether the solitary propagating wave is a crest or a trough, it is known as a bright or a dark soliton respectively. 

In cold atoms, experimental studies have been done on dark solitons as a dip in the density profile of a Bose-Einstein condensate~\cite{Denschlag07012000,PhysRevLett.83.5198,1751-8121-43-21-213001}, as well as systems of dark-bright soliton pairs~\cite{1367-2630-15-11-113028,PhysRevA.84.053630}. On the contrary, studies of bright matter wave solitons have been relatively few, despite offering similarly nuanced and interesting physics to dark solitons.  Early work observed the break-up of attractive $^7$Li and $^{85}$Rb condensates into soliton trains~\cite{Truscott_bright,1367-2630-5-1-373,PhysRevLett.96.170401}.  A single $6000$-atom $^7$Li bright soliton in an optical waveguide was created in 2002~\cite{Khaykovich17052002} and recently, a pair of neighbouring $100$-atom $^7$Li bright solitons were formed in a magnetic waveguide~\cite{PhysRevLett.112.060401}. A $2000$-atom $^{85}$Rb soliton has been studied while colliding with a repulsive barrier~\cite{cornish}.  Among many possible applications, a bright-soliton-based matter-wave interferometer has been proposed as a method to test the fine details of atom surface interactions~\cite{PhysRevD.68.124021}, and soliton collisions have been suggested as a mechanism for creating Bell-type entangled states~\cite{PhysRevLett.111.100406}.  Additionally, bright-solitonic atom interferometers hold great promise for precision measurements\,\cite{PhysRevA.88.053628,0953-4075-37-23-L02}, including measurements of gravity~\cite{RobinsAtomLaser,OurGravimeter,PhysRevA.88.053620,DebsBECgrav,BestAtomicGravimeter}, rotations and magnetic field gradients \cite{Russian,1367-2630-15-6-063006}, and tests of the weak equivalence principle~\cite{Tino2013203,Sorrentino, Schubert_arxiv,Aguilera_arxiv}. 

In this work, a bright solitonic matter wave of {${1\times10^4}$}~$^{85}$Rb atoms is produced in a horizontal optical waveguide.  Matter-wave propagation is investigated in the waveguide and the performance of a Mach-Zehnder matter-wave interferometer is compared across a range of $s$-wave scattering lengths.  It is found that there is a sharp optimum at negative $s$-wave scattering length that maximises interference fringe visibility in the interferometer.  The data indicates that a soliton-based interferometer dramatically outperforms its non-interacting counterpart, even at interferometer times as short as 1\,ms. Prospects for future experiments in this system are discussed with regards to both precision measurement and fundamental physics.

The experimental apparatus used here has previously been described in detail~\cite{Khun_arxiv,Hardman_arxiv}. Briefly, using hybrid magnetic and optical trapping~\cite{PhysRevA.86.063601,HybridTrap} and sympathetic cooling with $^{87}$Rb~\cite{altin:063103}, we produce samples of  $3.5\times10^{5}$~$^{85}$Rb at close to $1\,\mu$K in a crossed optical dipole trap. A magnetic bias field is then ramped on over 50\,ms to $\sim$140\,G, and then rapidly jumped through the 155\,G Feshbach resonance to 165.75\,G, zeroing the $s$-wave scattering of $^{85}$Rb and minimising three-body loss~\cite{PhysRevLett.85.728,PhysRevA.81.012713}.  Over a further $3.5$\,s the dipole trap intensity is smoothly ramped down, driving sympathetic evaporative cooling.  The final values give an optical trap with radial and axial trapping frequencies of $\omega_r=2\pi\times70(5)$\,Hz and $\omega_z=2\pi\times3(2)$\,Hz, respectively. Careful control of the sequence is required to ensure that there is no observable $^{87}$Rb coolant remaining at this point. In the last $0.5$\,s of this ramp the magnetic bias field is tuned to give a $^{85}$Rb $s$-wave scattering length of 300\,$a_0$, where the Bohr radius {$a_0=5.29\times10^{-11}\,$m}.  In this way, pure BECs of $1\times10^{4}$~$^{85}$Rb atoms are evaporatively formed in the crossed dipole trap.

The remaining atom cloud is transferred into a single beam dipole trap, effectively forming a waveguide for the atoms~\cite{PhysRevA.84.043618}.  The loading sequence follows our previous technique~\cite{PhysRevA.88.053620}, with the addition that, after forming the $^{85}$Rb condensate, the scattering length is ramped to $5\,a_0$ over 100\,ms, followed by a small ramp up in the the optical power of the waveguide over 200\,ms.  The weaker cross beam is then switched off suddenly, releasing the cloud into the guide, in which it is free to move longitudinally.   This is the initial condition for all waveguide experiments described in this letter. 
\begin{figure}[!b]
\centering{}
\includegraphics[width=1\columnwidth]{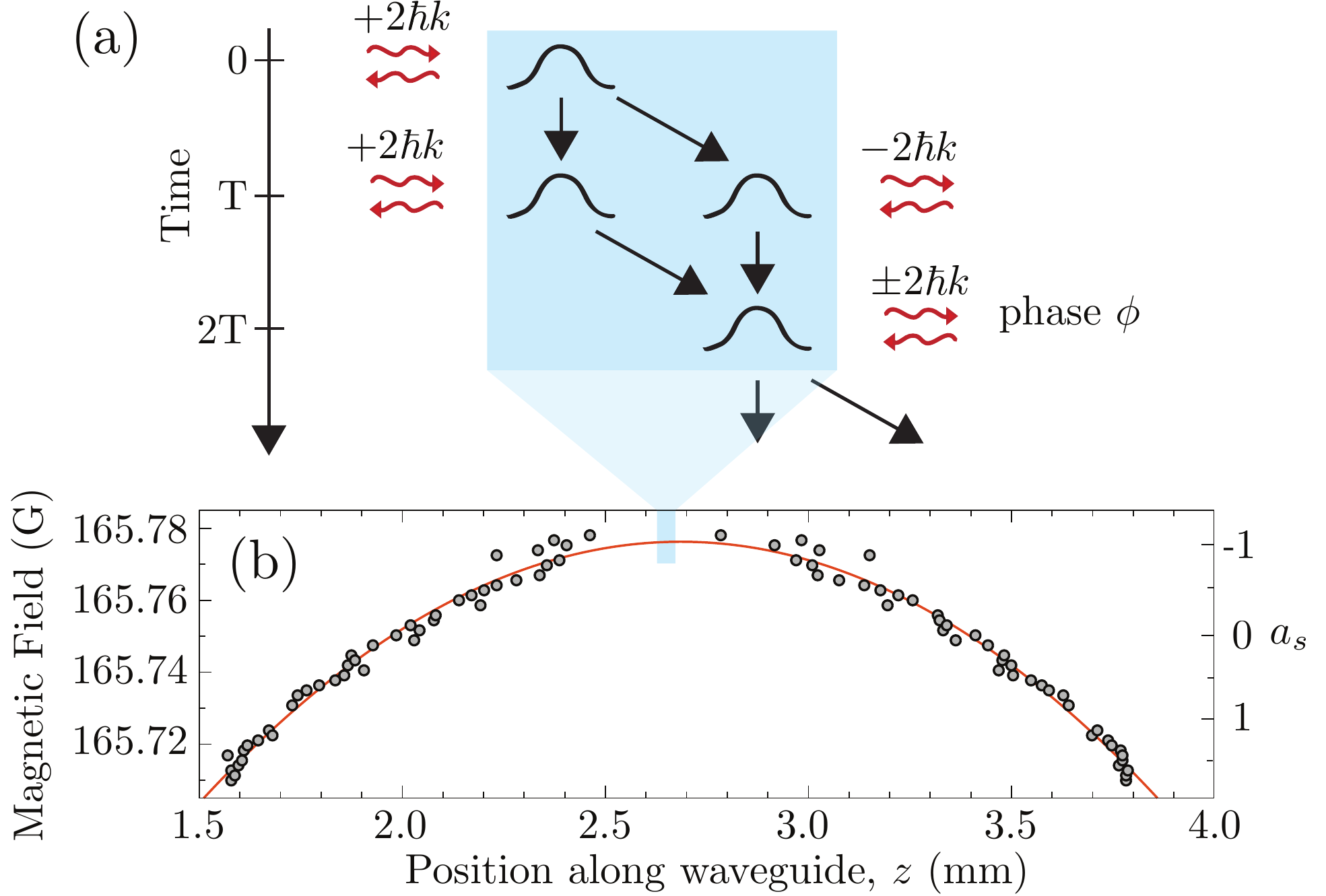}
\caption{(Color online) (a) Schematic of a Mach-Zehnder matter-wave interferometer constructed using optical Bragg transitions. (b) Measurement of the magnetic field curvature in the waveguide via r.f.~spectroscopy.  The magnetic field at each position is determined from the frequency required to drive {inter-$m_F$} transitions on an extended cloud of atoms in the guide. The red line is a parabolic fit to the data, indicating a repulsive harmonic potential with frequency $\omega_{z}=2\pi i\times3\text{\,Hz}$ along the waveguide. The interferometer and soliton expansion occurs in the region shaded in blue, slightly to the left of the maximum in the potential.}
\label{fig:mag}
\end{figure}

The Feshbach resonance used to manipulate the $s$-wave scattering length of the $\left|F=2,m_{\text{F}}=-2\right>$ state in $^{85}$Rb is controlled by an external magnetic bias field.  The resonance is well characterised by the equation ${a=a_{bg}\left(1-\frac{\Delta}{B-B_0}\right)}$, relating the $s$-wave scattering length, $a$, to the magnetic field, B, through the background scattering length, ${a_{bg}=-443\,a_0}$,  the width of the Feshbach resonance, {${\Delta=10.71}$\,G}, and the centre of the resonance, {${B_0=155.041}$\,G}~\cite{PhysRevLett.85.728,PhysRevA.81.012713}. Because the waveguide is extended in space, it is critically important to precisely characterise the bias field change along the waveguide. Radio-frequency (r.f.) transitions on an extended matter wave source are used to achieve this.  A 1\,$\mu$K sample of $^{87}$Rb atoms is released into the waveguide and allowed to expand over a second. A 10\,ms burst of r.f. couples the $\left|F=1,m_{\text{F}}=-1\right>$ and $\left|F=1,m_{\text{F}}=0\right>$ internal Zeeman states over a narrow frequency range according to the relation {${\hbar\omega_{\text{r.f.}}=\mu_{\text{B}}{\Delta}m_{\text{F}} g_{\text{F}}B}$}, where $\mu_{\text{B}}$ is the Bohr magneton, $g_{\text{F}}$ is the Land\'e $g$-factor for $^{87}\text{Rb}$ and the magnetic bias field is held at $B=165.776$\,G at the trap center. The bias field is then turned off and the resulting magnetic species are separated by a 2\,ms Stern-Gerlach pulse from the quadrupole coils. The locations at which each frequency couples the two internal states maps our magnetic field along the waveguide, and this is shown in Figure~\ref{fig:mag}~(b). A parabolic fit to this data yields a magnetic field curvature of {$\frac{\partial^2B}{\partial z^2}=-103(1)$\,mG/mm$^2$}. This curvature provides the dominant longitudinal potential for our $^{85}$Rb atoms which, according to the relation $\omega_z^2=\frac{\mu_B g_F m_F}{m_{\text{85Rb}}}\frac{\partial^2 B}{{\partial z}^2}$ (where $m_{85\text{Rb}}$ is the mass of $^{85}$Rb) gives an inverted harmonic potential along the waveguide with (anti) trapping frequency $\omega_{z}=2{\pi}i\times3\text{\,Hz}$ where $i=\sqrt{-1}$.  The experiments described here occur slightly to the left of the central region of the guide as indicated in blue in the figure, where we calculate a maximum magnetic field deviation due to both curvature and fluctuation of 4\,mG, allowing precise control of the $s$-wave scattering length of $^{85}$Rb. 

\begin{figure}[!htp]
\centering{}
\includegraphics[width=1\columnwidth]{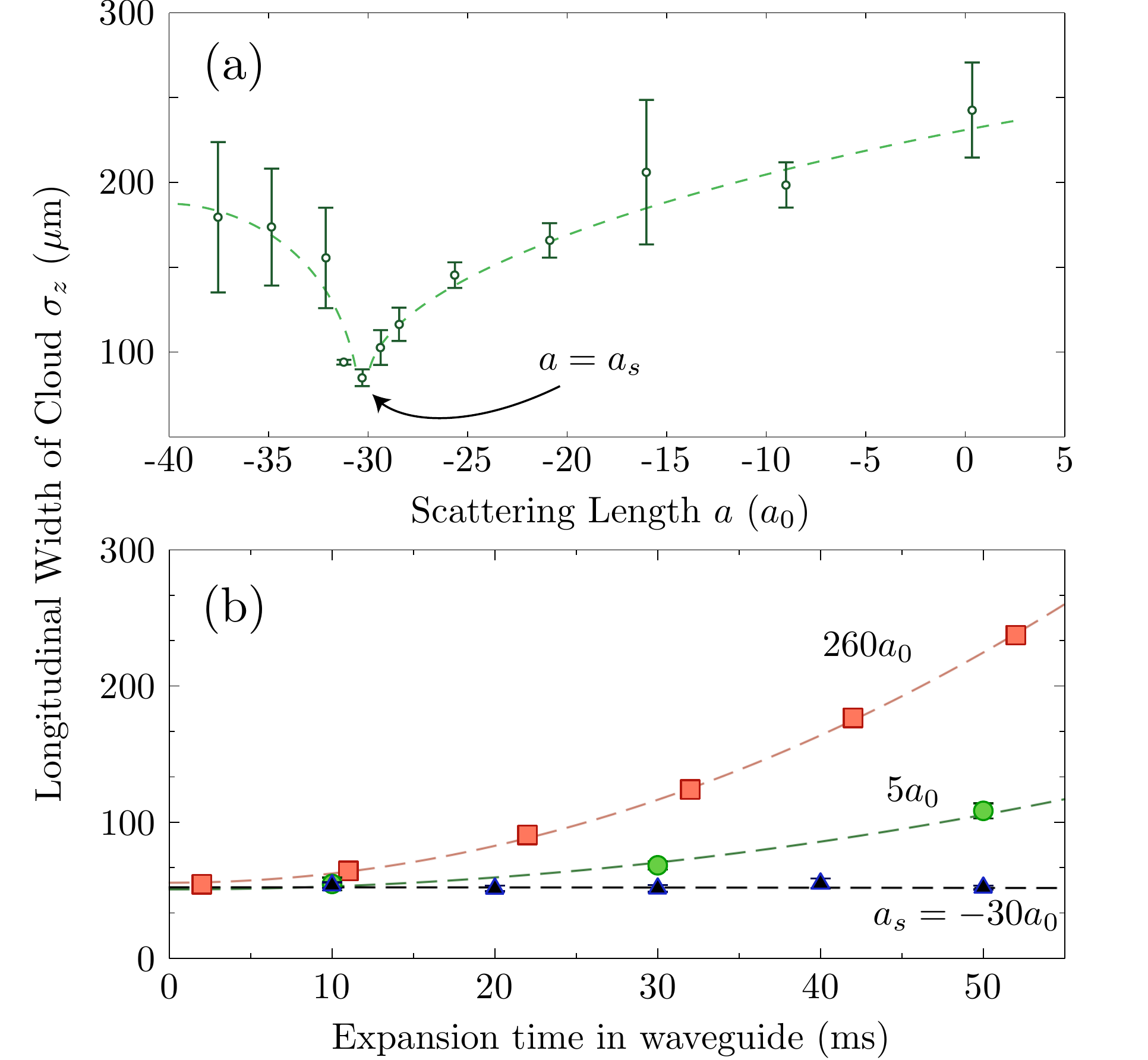}
\caption{(Color online) (a) Longitudinal width of a matter wave as a function of scattering length, measured after 90\,ms of free expansion in the guide. Green dashed lines are to guide the eye. The soliton parameter of $a_s=-30a_0$ is seen to minimise this expansion. (b) Comparison of longitudinal expansion along the guide for three different scattering lengths: the repulsive self-interaction of $a=260a_0$, the low interaction case of $a=5a_0$ and the soliton parameter of $a_s=-30a_0$. The dashed lines are parabolic fits to extract the acceleration of cloud width. For $a=a_s$ this acceleration is consistent with zero. Error bars shown in both (a) and (b) are statistical.}
\label{fig:expansion}
\end{figure}

\begin{figure*}[!htb]
\centering{}
\includegraphics[width=2\columnwidth]{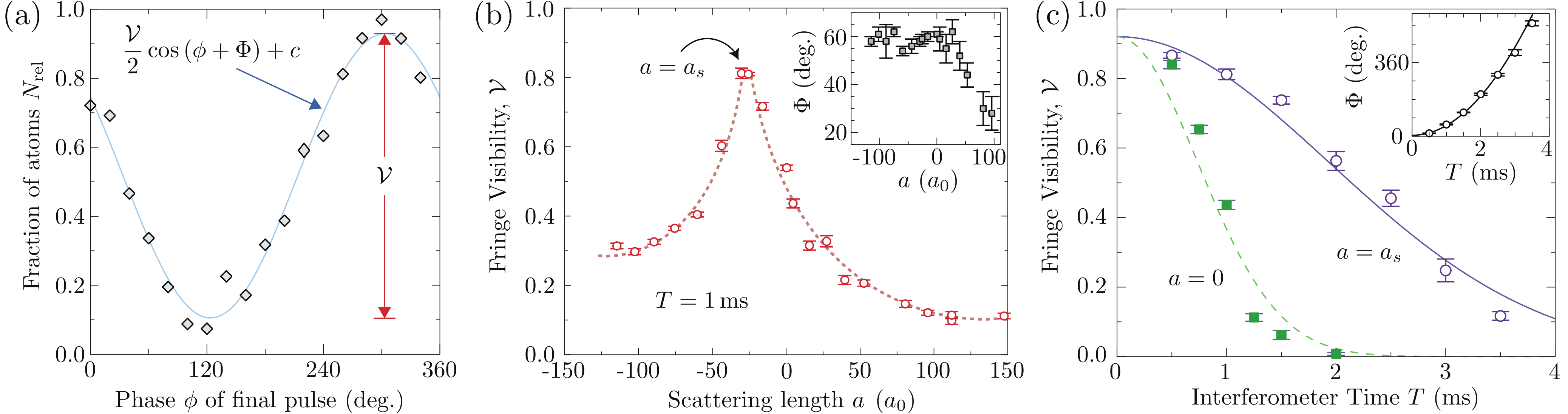}
\caption{(Color online) (a) Each interferometric fringe is fitted to a function $N_{\text{rel}}=(\mathcal{V}/2)\cos(\phi+\Phi)+c$ to extract the fringe visibility $\mathcal{V}$ and interferometric phase $\Phi$. The data shown here is the interference fringe for a soliton, with $a=a_s$. (b) Fringe visibility (open circles) as a function of $s$-wave scattering length $a$ for a {$T=1$~ms}  Mach-Zehnder  atom interferometer. Red dashed lines to guide the eye.  A dramatic increase in the fringe visibility is seen at the soliton parameter $a_s\approx-30a_0$ for our system. Inset: The interferometric phase $\Phi$ for various $a$. (c) Fringe visibility as a function of the interferometer time, $T$. Results for a non-interacting interferometer with $a=0$ (filled green squares), and the soliton interferometer with $a=a_s=-30a_0$ (open circles) are shown. Gaussian fits provide half-maximum decay times of 0.9~ms (green dashed line) and 2.3~ms (solid purple line) respectively. The solitonic matter-wave optimises the fringe visibility at all times we have tested. Inset: Measured interferometric phase $\Phi$ for the soliton interferometer as a function of $T$, along with a quadratic fit. All uncertainties in (b) and (c) are 1-$\sigma$ confidence intervals of the fit to each interferometric fringe.}
\label{fig:visibility}
\end{figure*}

After the $^{85}$Rb BEC is released into the waveguide, the scattering length is jumped to a value $a$ in order to study the expansion of the cloud over time. In Figure~\ref{fig:expansion}~(a) the longitudinal width of the guided matter wave after 90\,ms of propagation in the waveguide is plotted as a function of scattering length $a$.  There is a clear, sharp minimum in width of the expanded cloud at $a_s=-30a_0$, which we denote as $a_s$, the `soliton parameter' for our system. In Figure~\ref{fig:expansion}~(b) we compare the longitudinal expansion of a weakly-interacting matter wave (where $a=5a_0$, green circles)~\cite{PhysRevLett.100.080404,PhysRevLett.100.080405}, a repulsive matter-wave with $a=260a_0$ (red squares) and a condensate with attractive interactions set to the soliton parameter of $a_s=-30\,a_0$ (blue triangles) clearly showing the absence of dispersion, characteristic of a solitonic matter wave~\cite{PhysRevA.66.063602}. The dashed lines are parabolic fits to extract the acceleration of the cloud width $\frac{d^2\sigma_z}{dt^2}$ for each scattering length, which were found to be 67.5(6)\,mm/s$^2$, 22(1)\,mm/s$^2$ and -0.2(5)\,mm/s$^2$  respectively for the $260a_0$, $5a_0$ and $-30a_0$ clouds in this repulsive potential. Previous experimental realisations of solitons in the quasi-1D regime~\cite{Truscott_bright,1367-2630-5-1-373,PhysRevLett.96.170401,Khaykovich17052002,PhysRevLett.112.060401} in which the interaction parameter $|\alpha|\equiv N|a|\sqrt{m\omega_r/\hbar}<1$, differ from this work where $|\alpha|\approx 12\pm 2$, which is in the 3D regime. Solitons have previously been observed in a weak axial trap in the 3D regime for $|\alpha|\approx 3$ \cite{cornish}. The existence of a soliton in the parameter regime of this system in the presence of a weakly repulsive potential is consistent with previous theoretical calculations~\cite{PhysRevA.66.063602} as detailed in the supplementary material. This is supported by numerical simulation of both the 1D and 3D Gross-Pitaevski equations~\cite{PhysRevA.84.033632} confirming that dispersionless matter-wave propagation is possible in this regime. 

In order to investigate the properties of the soliton in this system, a Mach-Zehnder atom interferometer based on Bragg transitions is applied to the guided matter wave. The optical lattice used to effect Bragg transitions consists of up to 50\,mW in each of two counter-propagating beams, precisely aligned to be collinear with the waveguide optical trap by means of dichroic mirrors~\cite{PhysRevA.87.013632,FasterScaling}.  The lattice beams are collimated with a full width of 1.85~mm and detuned $\sim$100\,GHz to the blue from the D2 ${\left|F=2\right>}\rightarrow{\left|F'=3\right>}$ transition of  $^{85}$Rb. Arbitrary, independent control of the frequency, phase, and amplitude of each beam is achieved using a two-channel direct digital synthesizer (DDS) driving two separate AOMs. We use a standard three pulse $\pi/2\,-\pi\,-\pi/2$ configuration with a total separation time of $2T$, as illustrated in Figure~\ref{fig:mag}~(a). The first $\pi/2$ pulse splits the matter wave into two momentum states separated by $\Delta{p}=2\hbar{k}$. A time $T$ later, we apply a $\pi$ pulse to swap the momenta of the two states, allowing them to re-converge. After a further time $T$ the two momentum states are overlapped again and we recombine them with a final $\pi/2$ pulse.  After allowing some time for the clouds to physically separate, atom numbers $N_1$ and $N_2$ in the momentum state output ports  $0\hbar{k}$ and 2$\hbar{k}$ of the interferometer are measured using absorption imaging, and the relative fraction of atoms in the 0$\hbar{k}$ output state $N_{\text{rel}}=N_1/(N_1+N_2)$ is calculated for each run.   Interferometric fringes in $N_{\text{rel}}$ are obtained by scanning the optical phase $\phi$ of the final Bragg $\pi/2$ pulse using the DDS. An example fringe is show in Fig.~\ref{fig:visibility}~(a). A function $N_{\text{rel}}=(\mathcal{V}/2)\cos(\phi+\Phi)+c$ is fit to obtain the visibility $\mathcal{V}$ and interferometric phase $\Phi$ of the fringe. For the interferometer times presented here there is significant spatial overlap of the density of the two momentum states for the duration of the interferometer.

Figure~\ref{fig:visibility}~(b) shows the fringe visibility of a $T=1$~ms Mach-Zehnder interferometer as a function of the $s$-wave scattering length $a$ during the interferometer sequence.  The scattering length was abruptly changed from zero 0.4\,ms before the first $\pi/2$ pulse and back to zero 0.4\,ms after the last $\pi/2$ pulse. No significant systematic loss of atoms is measured over the range of $s$-wave scattering length observed. The sharp peak in the interferometer visibility occurs exactly at $a_s$, the `soliton parameter' identified from the expansion data in Figure~\ref{fig:expansion}, which therefore shows a solitonic matter wave very clearly outperforming a non-interacting cloud.  We attribute the increase in fringe visibility seen around $a=a_s$ to the lack of spatial dispersion as seen in Fig.~\ref{fig:expansion}~(b). Reduced longitudinal momentum width has been shown to increase visibility in atom interferometers in general  due to the frequency dependance of the Bragg transition~\cite{1367-2630-14-2-023009} and in our system in particular, reduced spatial dispersion has also been shown to increase mode-matching and therefore fringe visibility in the context of delta-kick cooling (Fig.~3(e) of Ref.~\cite{PhysRevA.88.053620}). Here, a more striking visibility peak due to the jump in scattering length from zero to the s-wave soliton parameter is also due to effectively freezing out the matter wave dispersion during the interferometer, increasing mode-matching in both position and momentum by conserving the phase-space density of each atom cloud. We hypothesise that this causes the interference to be more robust against visibility degradation due to any spatial inhomogeneity of the confining potential. Fringe visibility enhancement is also predicted for a solitonic interferometer due to collisional many-body entanglement~\cite{NoonContrastEnhancement}, which has already been demonstrated for the case of an optical soliton interferometer~\cite{PhysRevA.64.031801}.

To verify that the soliton parameter $a_s$ optimises fringe visibility for all $T$, the data in Figure~\ref{fig:visibility}~(b) was retaken for different interferometer times $T$. At all measured values of the interferometer time $T$, the system's `soliton parameter' that optimises fringe visibility has remained constant at $a_s=-30a_0$. In Figure~\ref{fig:visibility}~(c), the fringe visibility as a function of $T$ is plotted for both a non-interacting BEC with $a=0$ and a soliton with $a=a_s$. The coherence of the soliton interferometer has a half-maximum decay time $\sim2.5$ times as long as the non-interacting interferometer (as measured by a gaussian first-order coherence function~\cite{PhysRevA.59.4595,Hardman_arxiv}), again showing the clear advantage afforded by the solitonic matter wave. The inset of Fig.~\ref{fig:visibility}~(c) shows the interferometric phase measured as a function of interferometer time $T$ for the soliton interferometer. The repulsion observed previously \cite{Truscott_bright,1367-2630-5-1-373,PhysRevLett.89.200404} between solitons with a phase difference of $\pi$ is not seen here. Ref. \cite{0953-4075-41-4-045303} suggests that this behaviour is only observed for low relative velocities ($\Delta v<1$\,mm/s) whereas in our interferometer $\Delta v=2\hbar k/m_{\text{85Rb}}=12$\,mm/s.  A quadratic fit to the phase data is consistent with the cloud accelerating at $\left|\vec{a}\right|=5.2(1)\times10^{-2}\,\text{m/s}^{-2}$ due to both a slight tilt in the optical waveguide potential and the magnetic field gradient at the position of the atoms. The phase shift is quadratic because a Mach-Zehnder atom interferometer with $2\hbar\vec{k}$ momentum separation is sensitive to external acceleration $\vec{a}$ according to $\Phi=2\vec{k}\cdot\vec{a}T^2$~\cite{PhysRevA.88.053620,FasterScaling}. As the interferometer only samples a small region of the potential in Fig. \ref{fig:mag} (b) the center-of-mass acceleration is to a good approximation constant.

In conclusion, we have demonstrated that a solitonic matter-wave optimises the performance of a Mach Zehnder atom interferometer. 
A sharp optimum is evident in the peak in visibility of the interferometer fringes.
This new system offers an intriguing array of both fundamental and applied future research directions.  
Studies of soliton collision dynamics in a system with an interferometric probe offer the possibility to look for many-body entanglement~\cite{NoonContrastEnhancement}. 
It will also be possible to look for breather solitons~\cite{Matuszewski08112005} and yet more complicated soliton-like oscillations~\cite{Cardoso20102640}. 
Applications to precision measurement will require an in-depth study of the phase evolution of the solitons as a function of density and scattering length. 
The experiment is also suitable for studying polaritonic solitons, in which a BEC with zero $s$-wave scattering length can be made to self-interact via an applied optical lattice~\cite{PhysRevLett.110.250401}.

\section{Acknowledgements}
The authors gratefully acknowledge the support of the Australian Research Council Discovery program. C.C.N. Kuhn would like to acknowledge financial support from CNPq (Conselho Nacional de Desenvolvimento Cientifico e Tecnologico). J.E. Debs would like to acknowledge financial support from the IC postdoctoral fellowship program. 

\section{Supplementary Material}
\renewcommand{\theequation}{S\arabic{equation}}
\setcounter{figure}{0}
\renewcommand{\thefigure}{S\arabic{figure}}

In Reference~\cite{PhysRevA.66.063602}, a trial wavefunction

\begin{eqnarray}
\psi(\rho,z)=\frac{1}{\sqrt{2\pi l_\rho^2l_z}}\exp\left({-\frac{\rho^2}{2l_\rho^2}}\right)\text{sech}\left({\frac{z}{l_z}}\right)
\label{eq:trial}
\end{eqnarray}

was put in the Gross-Pitaevski energy functional to obtain the following energy surface with respect to both the axial and radial size of the condensate,
\begin{eqnarray}
\epsilon=\frac{1}{2\gamma_\rho^2}+\frac{\gamma_\rho^2}{2}+\frac{1}{6\gamma_z^2}+\frac{\pi^2}{24}\lambda^2\gamma_z^2+\frac{\alpha}{3\gamma_\rho^2\gamma_z}
\label{eq:energysurface}
\end{eqnarray}
where all the variables have been rescaled to the radial harmonic oscillator frequency $\omega_\rho$ or harmonic oscillator length $\sigma_\rho=\sqrt{\frac{\hbar}{m\omega_\rho}}$, i.e. energy $\epsilon=\frac{E}{\hbar \omega_\rho}$, radial width $\gamma_\rho=\frac{l_\rho}{\sigma_\rho}$, axial width  $\gamma_z=\frac{l_z}{\sigma_\rho}$, trap aspect ratio $\lambda=\frac{\omega_z}{\omega_\rho}=\frac{i\left|\omega_z\right|}{\omega_\rho}$ and interaction parameter $\alpha=\frac{Na}{\sigma_\rho}$.

\begin{figure}[!htp]
\centering{}
\includegraphics[width=0.7\columnwidth]{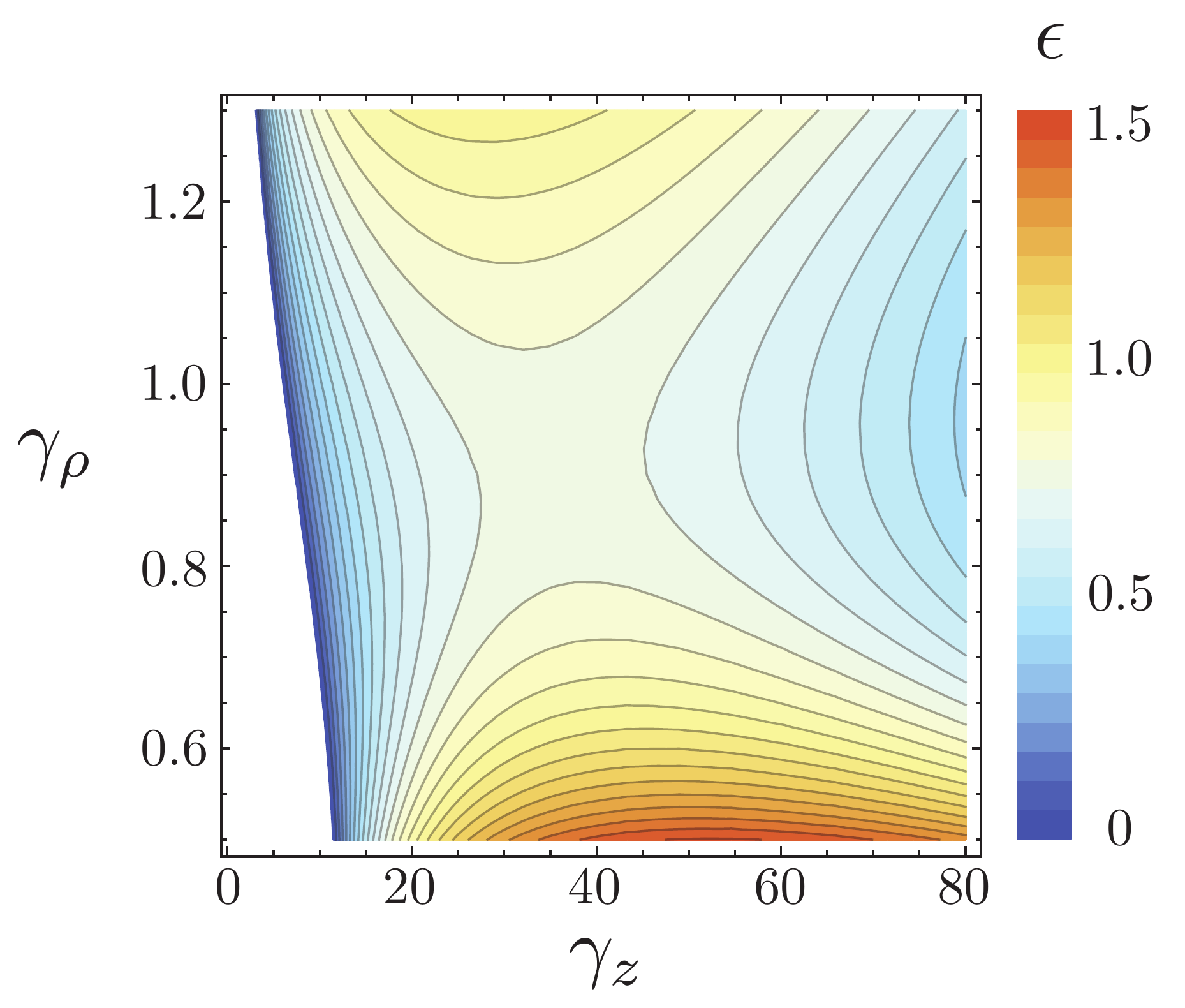}
\caption{(Color online) The energy surface defined by Equation (\ref{eq:energysurface}) for the parameters of $\omega_\rho=2\pi\times70$\,Hz, $\omega_z=2\pi i\times1$\,Hz, $N=1.5\times10^4$ atoms, $a=-30a_0$, showing a saddle point at $\gamma_\rho\approx0.9$ and $\gamma_z\approx40$, corresponding to $l_z\approx50\mu$m, the longitudinal width of our soliton.}
\label{fig:saddle}
\end{figure}

Soliton solutions are found at stationary points {(${\nabla\epsilon=0}$)} on this energy surface. Ref.~\cite{PhysRevA.66.063602} focussed upon the stable stationary point found at a minima of the energy surface, which only exists for a very limited parameter range (see Fig. 1 of that paper). Our experiment has a much larger interaction parameter $|\alpha|\approx12$ than anything in this region of stability (which is bounded by $|\alpha|\lesssim0.8$). In fact, by looking at the energy surface of Eq.~(\ref{eq:energysurface}) for parameters similar to those used in this experiment, we see that we are at a different stationary point, a saddle point, shown in the centre of Fig.~\ref{fig:saddle}. (c.f. Fig. 2 of Ref.~\cite{PhysRevA.66.063602}).

The exact values of the parameters should not be given too much weight, as their accuracy depends upon the accuracy of the trial wavefunction in Eq.~(\ref{eq:trial}). The real test is either an experiment such as the one presented in the main paper, or a full 3D simulation. Despite the instability of the saddle point soliton to perturbations, we still observe dispersionless propagation for over 50ms as shown in Figure 2(b) of the main paper.

\bibliographystyle{apsrev_v2}
\bibliography{refbibdesk}

\end{document}